# Technology Needs for Detecting Life Beyond the Solar System: A White Paper in Support of the Astrobiology Science Strategy

A Whitepaper in support of the Astrobiology Science Strategy

Authors:

Nick Siegler, Jet Propulsion Laboratory / California Institute of Technology;
nsiegler@jpl.nasa.gov
Matthew Bolcar, NASA Goddard Space Flight Center
Brendan Crill, Jet Propulsion Laboratory / California Institute of Technology
Shawn Domagal-Goldman, NASA Goddard Space Flight Center
Eric Mamajek, Jet Propulsion Laboratory / California Institute of Technology
Karl Stapelfeldt, Jet Propulsion Laboratory / California Institute of Technology

Abstract:

In support of the Astrobiology Science Strategy, this whitepaper outlines some key technology challenges pertaining to the remote search for life in exoplanetary systems. Finding evidence for life on rocky planets outside of our solar system requires new technical capabilities for the key measurements of spectral signatures of biosignature gases, and of planetary mass measurement. Spectra of Earth-like planets can be directly measured in reflected stellar light in the visible band or near-infrared using a factor 1e-10 starlight suppression with occulters, either internal (coronagraph) or external (starshade). Absorption and emission (reflected and thermal) spectra can be obtained in the mid-infrared of rocky planets transiting M-dwarfs via spectroscopy of the transit and secondary eclipse, respectively. Mass can be measured from the star's reflex motion, the reflex motion of a star, via either precision radial velocity methods or astrometry. Several technology gaps must be closed to provide astronomers the necessary capabilities to obtain these key measurements for small planets orbiting within the predicted temperate zones around nearby stars. These involved performance improvements, in some cases, 1-2 orders of magnitude from state-of-the-art or involve performances never demonstrated. The technologies advancing to close these gaps have been identified through the NASA Exoplanet Exploration Program's annual Technology Selection and Prioritization Process in collaboration with the larger exoplanet science and technology community.

## I. Whitepaper Objective

In support of the Astrobiology Science Strategy, this whitepaper will attempt to outline the key technology challenges pertaining to the remote search for life in extrasolar planetary systems.

## II. Science Questions

Thanks to NASA's Kepler space telescope, we now know that the Galaxy is teeming with planets. There is, on average, at least one exoplanet per star, and the majority of stars should contain an orbiting planetary system. We have learned that Earth-sized and "super-Earth" sized planets (between 0.5 and 2.5 earth radii) are the most commonly-sized planets, and while there are varying estimates on the frequency of such planets in the habitable zones of Sun-like stars, there is agreement that they are not rare (e.g. Fulton et al. 2017, Belikov et al. 2017). Consequently, we can move beyond the question of whether there are Earth-sized planets in the habitable zones of other stars, and we can now begin to ask (and answer) whether any of these planets harbor life.

The evidence for life on an exoplanet will most likely not be derived from a single measurement or observation, but rather will stem from a set of several measurements. This is because there are so many potential "false positives" for life that need to be ruled out - non-biological processes leading to perceived biosignatures. This results in a general drive towards as complete characterization of the planet and its stellar environment as possible, including: the spectral type and energy distribution of the host star, particularly in the ultraviolet; an understanding of the star's flare rates and coronal mass ejection history; a full inventory of the major gases in the planet's atmosphere; the physical properties of the planet's atmosphere such as characterization of the planetary surface in terms of the presence/absence of oceans, continents, and photosynthetic or other pigments; the orbital properties of the planet, in particular its semi-major axis and eccentricity; the mass, radius, and density of the planet; and the orbital properties and masses of other planets in the system.

Interestingly, all of these measurements fall into four kinds of observations: spectra of the planet, photometry of the planet, mass of the planet, and spectra of the star. But with specific interest in mature rocky planets in stars' habitable zone, only spectral measurements of stars can be acquired by instruments within the current technology state-of-art (SOA); the other three require technology development.

## III. Key Technical Capabilities

To obtain the measurements listed in the previous section scientists need to develop the capabilities to do two very difficult things very well – 1) spectrally characterize the atmosphere and surface of Earth-like planets, and 2) measure their mass. These are what we refer to as desired key technical capabilities. Achieving these capabilities across a broad range of wavelengths, along with the existing capability to spectrally characterize stars, will allow astronomers to collect the necessary data.

There are three key technology areas requiring advancement to achieve these two capabilities:

- Direct imaging of exoplanets (so as to perform reflection/emission spectroscopy)
- Transit (absorption spectroscopy) / secondary eclipse (emission spectroscopy)
- Stellar reflex motion

## IV. Technology Gaps

NASA's Exoplanet Exploration Program (ExEP) identifies technology gaps pertaining to possible exoplanet missions and works with the community to identify and track technologies to prioritize for investment, and ultimately to close the gaps. These technologies are summarized in the ExEP's annually-updated Technology List (Crill & Siegler 2017a) and captured in detail in their Technology Plan Appendix (Crill & Siegler 2017b). A possible roadmap to mature these technologies is described in Crill & Siegler (2017c) The gaps in performance, as related to their technology areas, are:

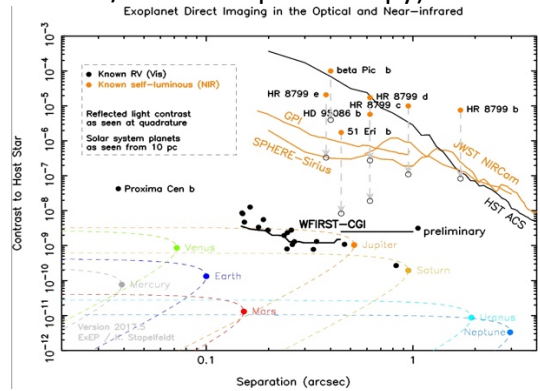

Figure 1: Contrast (ratio of planet brightness to host star brightness) versus apparent angular separation. For a larger version of this figure including a full caption go to exoplanets.nasa.gov/exep/technology/technology-overview/

### Direct imaging of exoplanets
**Starlight suppression for reflection (or emission) spectroscopy.** Suppression of starlight in order to bring orbiting exoplanets into view requires either starlight occultation or nulling. Those are the only two approaches known. Starlight occultation technologies include both internal (coronagraph) and external (starshade) approaches and have continued to progress this decade largely motivated by the 2010 Decadal Survey's number one medium-scale size recommendation - developing the technologies to enable the imaging of rocky planets in the next decade. NASA chose the simpler single room-temperature telescope observing at short wavelengths over the more complex multiple telescope, cryogenic, formation flying architecture that the long wavelength observations starlight nulling would have required.

Ground-based telescopes with coronagraphs, even next generation instruments on future 30 m-class telescopes, are expected to be fundamentally limited to $10^{-8}$ contrast sensitivities due to the residual uncorrected errors from atmospheric turbulence correction (Stapelfeldt 2005; Traub & Oppenheimer 2010). WFIRST's technology demonstration coronagraph will be the first high-contrast coronagraph in space possessing wavefront-correcting optics, such as a low-order wavefront sensor and deformable mirrors, to achieve contrast sensitivities between $10^{-8}$ and $10^{-9}$. WFIRST and its 2.4 m telescope is planned to launch in the mid-2020s. To observe an Earth-size exoplanet orbiting in the habitable zone of a Sun-like star, however, would require sensitivities to contrast ratios of $10^{-10}$ or better (see Fig 1); large super-Earths could be slightly more favorable. This is 1-2 orders of magnitude more demanding than WFIRST's expected performance and 2 orders more than future ground-based telescopes.

Coronagraphs with little to no central obscuration will have the highest likelihood to achieve the $10^{-10}$ contrast goal while simultaneously achieving high throughput. The Hybrid Lyot coronagraph achieved $6 \times 10^{-10}$ contrast at 10% bandwidth (Trauger et al. 2011). The Decadal Survey Testbed, an ExEP facility for testing next-generation coronagraphs, is being commissioned in the spring of 2018 to advance performance to better than $10^{-10}$. Future large space telescopes are very likely to have segmented apertures with secondary mirror obscurations. To address the challenges in achieving the contrast goals while maintaining high throughput, the ExEP Segmented Coronagraph Design & Analysis study was commissioned in 2016 to work with leading coronagraph designers. At the time of this writing there are about a couple candidates on the path to meet the requirements. If successful, the masks and optics for these designs will be fabricated and tested in multiple testbeds before the end of the decade. A coronagraph solution to imaging exo-Earths in their stars' habitable zone appears to be on track.

The starshade is currently being advanced under an ExEP technology development activity. While a full-scale starshade has never been demonstrated, a preliminary assessment (Seager et al. 2015) has developed design models predicting better than $10^{-10}$ contrast and a sub-scale validation demonstration is far along. However, to test the diffraction regime expected in space (i.e. flight Fresnel number) and operating within a practical-sized testbed (77 m), the demonstration is being conducted with only a 25 mm starshade (the separations between the "spacecraft" increase with the square of the starshade radius so testing large sizes require very large testbeds). To test the robustness of the optical models, intentions are to conduct additional suppression testing at longer wavelengths and more than one starshade size.

The scattering of Sun light off the starshade's petal edges is an important design factor and materials that are sufficiently thin, low-reflectivity, and malleable for stowage are being investigated. The starshade also requires a precise and stable structural deployment from a stowed configuration that is unique to previous NASA missions (< 1 mm petal positioning error). However, there does not appear to be any show-stoppers for a starshade to be mechanically designed to deploy to this tolerance. The starshade appears to be on a path to reach TRL 5 in the early part of the next decade and be ready for a potential rendezvous mission with WFIRST (pending recommendation by the 2020 Decadal Survey).

**Contrast stability.** Due to the extremely low rate of photons from distant exoplanets (in the range of about a photon per minute(s) in the case of the WFIRST coronagraph), achieving spectroscopy at a sufficient signal-to-noise ratio will require long integration times. The extreme starlight suppression must be maintained as the space observatory experiences drifts (both thermal and dynamic changes) during the integration. Large segmented telescopes will particularly be challenged by the need to achieve a stable back-structure and maintain a large number of individual segments as a single paraboloid. Lastly, spacecraft disturbances such as those initiated by reaction wheels must be dampened before reaching the coronagraph.

In the case of coronagraphy, error budgets for wavefront error stability range typically between 10-100 pm rms for a telescope and instrument system (Nemati et al 2017). This is 1-2 orders of magnitude more demanding than what has been demonstrated in space or in the lab. On-going analyses being conducted by the HabEx and LUVOIR study design teams will best determine the likelihood of these telescope systems meeting the very demanding wavefront error stability requirements.

In the case of a starshade-only mission, telescope stability requirements are significantly looser and do not exceed the SOA. Solutions for sensing and alignment control between the two spacecrafts have been developed and subscale demonstrations are being conducted in the lab.

**Detection sensitivity.** Even after suppressing the starlight to achieve the demanding contrast sensitivities and maintaining the required wavefront error stability, the light from the exoplanets must still be detected. The low flux of the targets requires a detector with read noise and spurious photon count rate as close to zero as possible, and that maintains adequate performance in the space environment. The SOA is dependent on the wavelength band but detectors must perform at or near the photon counting limit in the near-UV, the visible band, the NIR, and the MIR. Across this wavelength range, the SOA detectors are semiconductor-based devices. WFIRST's electron multiplying charge coupled device (EMCCD) detectors have achieved adequate noise performance in the visible band, though longer lifetime in the space radiation environment is desirable. Similar EMCCD devices, with delta doping, may already have adequate performance in the near-UV. HgCdTe detectors are the SOA in the NIR. JWST/MIRI's detectors are expected to establish the SOA in MIR detection sensitivity, and future direct imaging is likely to require detectors that exceed it. It is likely that the detection sensitivity gap can be closed in the next decade, as a range of choices are close to meeting the requirements.

**Angular resolution and collecting area**.  Large space telescopes offer many benefits in the determination of exoplanet habitability. Improved spatial resolution allows for a larger exoplanet yield, particularly those in the habitable zones of nearby stars. The larger collecting area also enables higher spectral resolution to better define molecular features as well as overall improved detection sensitivity. A larger telescope also better rejects the extended diffuse brightness of exozodiacal light that could obscure exoplanets. The largest monoliths flown in space are the 2.4 m Hubble Space Telescope, optimized for visible and UV astronomy, and Herschel's 3.5 m telescope, optimized for the far-IR. The James Webb Space Telescope will establish the SOA in space telescopes with a 6.5 m primary mirror made up of 18 co-phased hexagonal beryllium segments. Current mission concept studies range from 4 m monoliths to 15 m segmented telescopes.

Large glass monoliths have been commonly fabricated for ground-based telescopes. If future heavy-lift launch vehicles like the Space Launch System become a reality then the opportunity for a 4 m-class monolith becomes a possibility. Large monoliths will advance exoplanet science but will not directly lead to subsequent larger telescope architectures (> 10 m). One-meter class silicon carbide and glass segmented mirrors have fabrication heritage and appear to be promising options if the design teams can show there is sufficient control authority to meet the contrast goals.

## Transit/secondary eclipse spectroscopy
**Spectroscopic Sensitivity**.  To enable precise transit or secondary eclipse spectroscopy, the detector response must exhibit photometric stability over the time scales of a transit, typically hours to days. Spitzer/IRAC has achieved photometric stability of order 60 parts per million on transit time scales. JWST/MIRI is expected to achieve stability between 10-100 ppm. A stability of 5-10 ppm in the mid-IR is needed in order to measure the atmospheres of Earth-sized planets transiting nearby M-dwarfs. Astrophysical limits to this technique due to stellar activity need to be quantified.

The path to close the technology gap in transit spectroscopy is currently not known. First, astrophysical limits should be examined further to find likely fundamental limits to stellar stability. The sources of instability in detector/telescope systems must be studied to determine where future technology investments will be most effective. Photometric instabilities of a mid-IR detector system may be driven by fundamental detector materials properties, cryogenic detector readout circuitry, or other instabilities in the system. This should be done along with modeling the on-orbit calibration, which will mitigate the detector requirements to some level.

Valuable lessons will be learned from performing these measurements with JWST/MIRI in the early 2020s.

**Stellar reflex motion**
**Radial stellar motion sensitivity.** Radial velocity (RV) measurements of the reflex motion of a star can be a way to infer the minimum mass and orbital parameters of planets orbiting the star. The HARPS instrument has recently achieved 40 cm/s precision (Feng et al. 2017). The next generation of ground-based RV instruments coming online in the next 1-2 years are expected to achieve 20-30 cm/s instrumental sensitivity per measurement. The reflex motion of a Solar-mass star due to an orbiting Earth-mass planet at 1 AU is ~10 cm/s over 1 year, and both measurement and systematic errors must be kept below that.

The biggest uncertainty in closing this gap is understanding the astrophysical limits due to natural stellar jitter. At this point the path forward to achieving 1 cm/s sensitivity and closing the gap is unclear but may be better understood upon completion of NASA-chartered probe study, and through experience at mitigating systematics errors in ground-based RV instruments measurements.

**Tangential Stellar Motion Sensitivity.** By performing sensitive astrometry of a star over time, the mass and orbital parameters of orbiting exoplanets can be measured. GAIA's initial data release achieved a typical 300 microarcsecond position error, but GAIA is expected to achieve 10 microarcsecond sensitivity in the positions of many stars in subsequent data releases, sensitive enough to reveal many Jupiter-mass exoplanets. A precision of 0.3 microarcsecond per measurement is needed in order to enable the detection of Earth-mass planets at a distance of 10 pc.

The path to closing this technology gap in astrometry is not clear. It is possible that astrophysical limits due to variable stellar surface structure may prevent astronomers from reaching this precision. The inherent instabilities of stars needs further understanding and sources of instrument instability and the ability to calibrate them using techniques such as interferences fringes or diffractive pupils should be modeled.

## V. Conclusion

The existing technology gaps needing to be bridged to provide astronomers the necessary capabilities to obtain the key measurements are, in some cases, 1-2 orders of magnitude from the SOA or involve performances never demonstrated. The technologies being developed to close these gaps have been identified and are being advanced. They are currently at various degrees of readiness. These technologies are summarized in the ExEP's annually-updated Technology List (Crill & Siegler 2017a) and captured in detail in their Technology Plan Appendix (Crill & Siegler 2017b). A life-finding large mission recommendation by the 2020 Decadal Survey would be required to prioritize, focus, and accelerate technology development in the next decade to enable a launch in the 2030s.

----
References can be found at the following URL:
https://exoplanets.nasa.gov/internal_resources/774_References_for_Astrobiology_Technology_Whitepaper.pdf